\documentclass[prl,showpacs,twocolumn]{revtex4}

\newcommand {\etal}{{\it et al.}}

\newcommand {\CFO}{CuFeO$_2$}
\newcommand {\CFAO}{CuFe$_{1-x}$Al$_{x}$O$_2$}
\newcommand {\CFGO}{CuFe$_{1-x}$Ga$_{x}$O$_2$}

\newcommand {\Pin}{$P_{[110]}$}
\newcommand {\Hout}{$H_{[001]}$}
\newcommand {\Hin}{$H_{[110]}$}
\newcommand {\Hbar}{$H_{[1\bar{1}0]}$}
\newcommand {\Pbar}{$P_{[1\bar{1}0]}$}

\newcommand {\AngleH}{$\theta_H$}
\newcommand {\AngleP}{$\theta_P$}

\usepackage{graphicx}
\usepackage{bm}
\usepackage{pifont}
\usepackage{amsmath}

\begin{document}

\title{Magnetic digital flop of ferroelectric domain with fixed spin chirality in a triangular lattice helimagnet}

\author{S. Seki$^1$, H. Murakawa$^2$, Y. Onose$^{1,2}$, and Y. Tokura$^{1,2,3}$} 
\affiliation{$^1$ Department of Applied Physics, University of Tokyo, Tokyo 113-8656, Japan \\ $^2$  Multiferroics Project, ERATO, Japan Science and Technology Agency (JST), Tokyo 113-8656, Japan \\ $^3$ Cross-Correlated Materials Research Group (CMRG), RIKEN Advanced Science Institute, Wako 351-0198, Japan}

\date{November 13, 2009}

\begin{abstract}

Ferroelectric properties in magnetic fields of varying magnitude and direction have been investigated for a triangular-lattice helimagnet CuFe$_{1-x}$Ga$_{x}$O$_{2}$ ($x$=0.035). The magnetoelectric phase diagrams were deduced for magnetic fields along [001], [110], and [1$\bar{1}$0] direction, and the in-plane magnetic field was found to induce the rearrangement of six possible multiferroic domains. Upon every $60^\circ$-rotation of in-plane magnetic field around the $c$-axis, unique 120$^\circ$-flop of electric polarization occurs as a result of the switch of helical magnetic $q$-vector. The chirality of spin helix is always conserved upon the $q$-flop. The possible origin is discussed in the light of the stable structure of multiferroic domain wall.\end{abstract}
\maketitle

Magnetoelectric (ME) effect, i.e. magnetic (electric) induction of electric polarization $P$ (magnetization $M$), has long been investigated\cite{Review2}. The most promising approach to achieve gigantic ME effect is the magnetic and/or ferroelectric domain control in multiferroics, materials with both magnetic and ferroelectric (FE) order. In general, the formation of domain structure depends on the symmetry\cite{Schmid}; if each domain is characterized by both magnetic and dielectric order parameters, ME effects are obtained from modulation of domain distribution by electric field ($E$) or magnetic field ($H$). This strategy was first demonstrated on a ferroelectric weak ferromagnet (FM) Ni$_3$B$_7$O$_{13}$I with persistent 180$^\circ$-reversal of $P$-vector under 90$^\circ$-rotation of $H$\cite{Boracite}. However, this type of ME control has seldom been achieved, because of the rareness of similar FE/FM compounds\cite{Spaldin} and the strict symmetry restrictions for selective domain switching\cite{Schmid}.

One important breakthrough was achieved by the discovery of ferroelectricity of magnetic origin\cite{Review1}. In some frustrated magnets, spiral spin order is found to induce finite $P$, which is now explained in terms of the inverse effect of Dzyaloshinskii-Moriya (D-M) interaction\cite{Katsura}. This model predicts the coupling between the sign of $P$ and spin chirality (left-handed or right-handed spin rotation), which was experimentally confirmed by studies of polarized neutron scattering\cite{RMnO3_PolarizedNeutron, LiCu2O2_PolarizedNeutron}. In some helimagnets like TbMnO$_3$\cite{Kimura} and MnWO$_4$\cite{MnWO4}, a specific direction of static $H$ induces the 90$^\circ$-flop of spin-spiral plane and accordingly of $P$-direction. Some hexaferrites show smooth rotation of $P$-vector and spin-spiral plane under rotating $H$\cite{Plumbite, Ishiwata}. Likewise, the collinear spin order can also be coupled with $P$, due to symmetric exchange striction\cite{Review1}; TbMn$_2$O$_5$ with collinear spin order shows $H$-induced 180$^\circ$ reversal of $P$\cite{TbMn2O5}. In all these compounds, however, the $P$-direction change is not persistent after the removal of $H$, since directional change of $P$ originates from the $H$-induced modulation of spin structure.

To apply multiferroics to the novel high-density non-volatile storage device, such as an $H$-controlled ferroelectric memory, the persistent switch of $P$-direction as well as increase of the number of  switchable meta-stable states are highly desirable\cite{Memory}. While several approaches have realized the $H$-induced persistent change of $P$-direction\cite{Boracite, CoCr2O4, Rotation}, they are all limited to the $180^\circ$ switch between $\pm P$ states. Aside from the non-volatile nature, $H$-induced discontinuous switch of $P$-direction among more than two different axes (i.e. directional switch of $P$ by other than $90^\circ$ or $180^\circ$) has never been achieved. In this study, we extend the concept of domain switching to demonstrate the persistent magnetic control of six ferroelectric domains with different $P$-directions for triangular lattice ferroelectric helimagnet {\CFGO} ($x$=0.035). The flop of magnetic modulation vector can be induced upon every $60^\circ$-rotation of in-plane $H$ around the $c$-axis, which leads to every $120^\circ$-flop of $P$-direction within the triangular-lattice basal plane. Interestingly, the chirality of spin-spiral is always conserved upon the $P$-flop. We discuss the possible origin in the light of the stable structure of multiferroic domain wall.

\begin{figure}
\begin{center}
\includegraphics*[width=8.5cm]{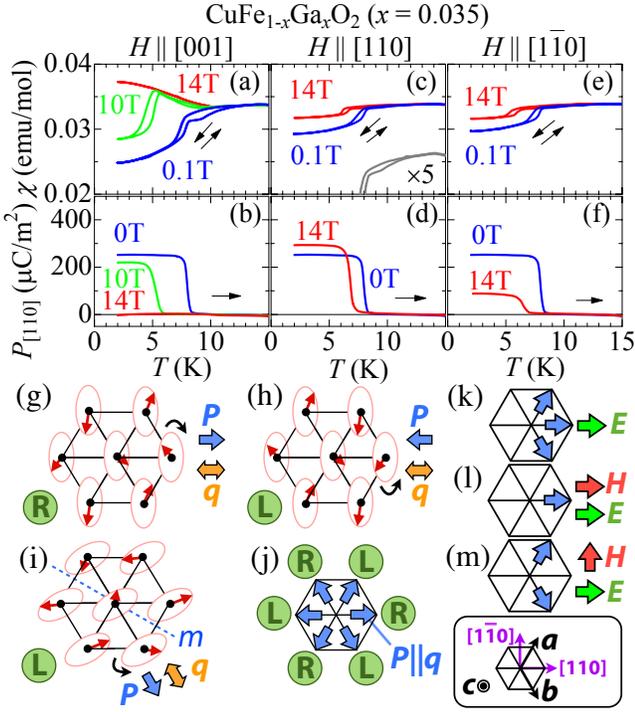}
\caption{(color online). Temperature ($T$) dependence of magnetic susceptibility ($\chi$) and [110] component of electric polarization ($P$). Magnetic field ($H$) is applied along (a),(b) [001], (c),(d) [110], and (e),(f) [1$\bar{1}$0], respectively. Arrows indicate the direction of $T$-scan. In (c), a magnified profile of $\chi$ at 0.1T (arbitrarily off-set) is also shown. (g)-(i) Three out of six possible multiferroic domains with proper screw magnetic structure on triangular lattice. Circled "R" and "L" denote the chirality of spin spiral. Directions of  $P$ and magnetic $q$-vector are also indicated. (j)-(m) Distribution of multiferroic domain(s) favored under various $H$ and electric field ($E$). The spin chirality corresponding to each $P$-domain is shown in (j).}
\end{center}
\end{figure}

{\CFO} is known as a member of triangular lattice antiferromagnets with spin frustration. Magnetic moment is carried by Fe${^{3+}}$ ion with $S=5/2$, and each element (Cu/Fe/O) forms triangular lattice respectively. They stack along the $c$-axis in order of Cu${^+}$-O${^{2-}}$-Fe${^{3+}}$-O${^{2-}}$-Cu${^+}$, and crystalize into the delafossite structure with centrosymmetric space group $R\bar{3}m$\cite{CuFeO2_FloatingZone}. {\CFO} undergoes subsequent magnetic phase transitions (CM-4 $\rightarrow$ NC $\rightarrow$ CM-5) under $H$ applied along the $c$-axis ({\Hout}), while keeping the in-plane component of magnetic modulation vector ($q$) parallel to $\langle 110 \rangle$ \cite{CuFeO2_PhaseDiagram, Notation}. Hereafter, we treat $q$ as the director (with no +/- sign). CM-4 and CM-5 have collinear spins along the $c$-axis, and correspond to $\uparrow \uparrow \downarrow \downarrow$ and $\uparrow \uparrow \uparrow \downarrow \downarrow$ spin structures, respectively\cite{CuFeO2_PhaseDiagram}. Ferroelectricity is observed only in the NC phase with proper screw magnetic structure\cite{CuFeO2_Polarization}, where spiral spin rotates within the plane perpendicular to the magnetic $q$-vector\cite{CuFeO2_ProperScrew}. The $P$ in NC(FE) appears parallel to $q$-vector (Fig. 1(g)), and the spin chirality is confirmed to be reversed for the reversal of $P$-direction (Fig. 1(h))\cite{CuFeO2_PolarizedNeutron,CuFeO2_Arima, CuFeO2_Comprehansive}. The present ferroelectricity with  $P \parallel q$ cannot be explained by the simple inverse D-M model, which predicts $P \perp q$\cite{Katsura}. Instead, the variation of $\pi$-bonding between Fe and O ions under the influence of spin-orbit interaction has been suggested as the origin of ME coupling\cite{CuFeO2_Arima, Jia1, CuFeO2_PolarizedNeutron}. The critical $H$ to induce the NC(FE) phase can be reduced to zero by nonmagnetic impurity doping on Fe$^{3+}$ site. This was originally shown for Al-doping but with much reduced $P$\cite{CuFeO2_Seki, CuFeO2_Al} due to strong pinning of domain wall\cite{CuFeO2_Comprehansive}.  This problem can be overcome by Ga-doping\cite{CuFeO2_Ga}, and here we adopt {\CFGO} ($x$=0.035) as a target multiferroic. Without external field, this compound has six equivalent ferroelectric domains with $P \parallel \langle 110 \rangle$ (Fig. 1(j)) due to high symmetry of underlying triangular lattice. From the $P \parallel q$ relationship, they correspond to six spiral magnetic domains with three different $q \parallel \langle 110 \rangle$ axes and two spin-chiral degrees of freedom.

\begin{figure}
\begin{center}
\includegraphics*[width=8.5cm]{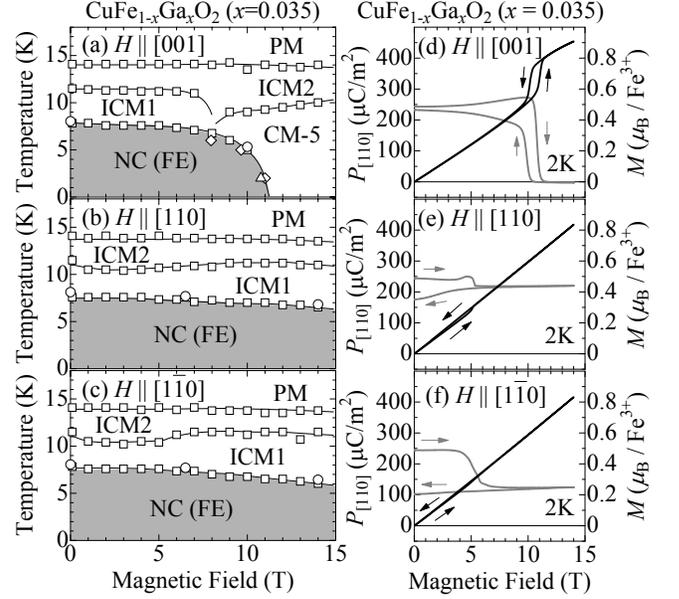}
\caption{$H$-$T$ phase diagrams with $H$ parallel to (a) [001], (b) [110], and (c) [1$\bar{1}$0] direction. Circles, triangles, squares, and diamonds are the data points obtained from $P$-$T$, $P$-$H$, $\chi$-$T$, and $M$-$H$ curves, respectively. All the data were taken from the increasing $T$ or $H$ runs\cite{Notation}. Ferroelectric (FE) state is observed in the shadowed region. (d)-(f) $H$-dependence of $P$ (gray line) parallel to [110] and magnetization ($M$: black line) at 2K, with $H$ applied along (d) [001], (e) [110], and (f) [1$\bar{1}$0] direction. In all measurements, $H$ was swept from 0T to 14T and then back to 0T, after $T$ was lowered to 2K at 0T. The arrows indicate the direction of $H$-scan.}
\end{center}
\end{figure}

Single crystals of {\CFGO} ($x$=0.035) were grown by a floating zone method\cite{CuFeO2_FloatingZone}. They were cut into a rectangular shape with the paired (110) and (1$\bar{1}$0) surfaces, on which silver paste was painted as electrodes. To deduce $P$, we measured the polarization current with constant rates of temperature($T$)-sweep (2K/min), $H$-sweep (80 Oe/sec), or $H$-rotation (1$^\circ$/sec), and integrated it with time. To enlarge the population of specific $P$-domains, the poling electric field ($E=250$kV/m) was applied along [110] in the cooling process and removed just prior to the measurements of polarization current. $M$ was measured with a SQUID magnetometer.

\begin{figure}
\begin{center}
\includegraphics*[width=8.5cm]{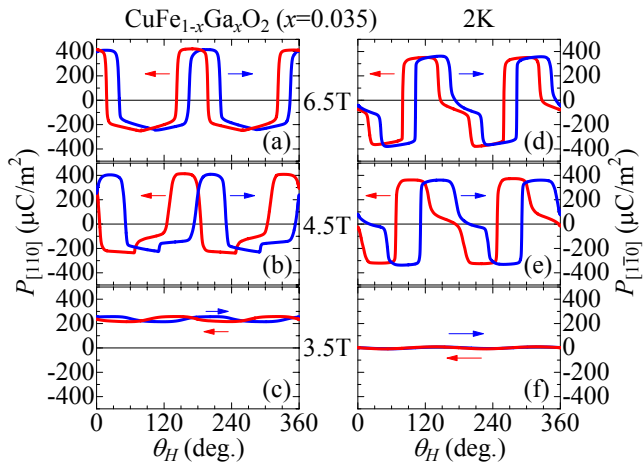}
\caption{(color online). (a)-(c) [110] and (d)-(f) [1$\bar{1}$0] components of $P$ simultaneously measured in $H$ rotating within the (001) plane. $\theta_H$ denotes the angle between $H$-vector and the [110] axis (see Fig. 4(d)). Arrows indicate the direction of $H$-rotation. Absolute value of $P$ was determined by $T$-scan. }
\end{center}
\end{figure}

Since only the properties under zero magnetic field have been reported so far for Ga-doped {\CFO}\cite{CuFeO2_Ga}, we first establish $H$-$T$ phase diagrams for the $x$=0.035 specimen in {\Hout}, {\Hin} and {\Hbar} (Figs. 2(a)-(c)). They are determined from the measurements of $T$- and $H$-dependence of $M$ and [110] component of $P$ ({\Pin}) (Figs. 1(a)-(f) and Figs. 2(d)-(f)), by analogy with the case for {\CFAO} ($x$=0.02) under {\Hout}\cite{CuFeO2_Seki, CuFeO2_Al}. Here, paramagnetic and two different sinusoidally-modulated collinear incommensurate magnetic phases\cite{Notation, CuFeO2_OPD} are referred to as PM, ICM1 and ICM2, respectively. With any direction of $H$, the boundary of the FE phase (shadowed region) always coincides with that of the NC magnetic phase, which ensures the coupling between ferroelectricity and proper-screw magnetic structure. In $T$-scan profiles, the onset of spiral magnetic order, coupled with the emergence of ferroelectric $P$, can be detected as the sudden drop of magnetic susceptibility $\chi (=M/H)$. While the NC(FE) phase is replaced by CM-5 under $H_{[001]}>12$T, we found that NC(FE) survives against in-plane $H$ ({\Hin} and {\Hbar}) up to 14T. 

Even with in-plane $H$, the magnitude of {\Pin} shows significant $H$-dependence, as shown in Figs. 1(d) and (f): {\Pin} increases with {\Hin} and decreases with {\Hbar}. This behavior can be interpreted as the rearrangement of multiferroic domains, as described below. In general, antiferromagnetically ordered spin moments prefer to lie within the plane perpendicular to $H$, as typically observed for the spin flop transition. Thus, in case of proper screw magnetic structure, $H$ favors the magnetic domain with $q \parallel H$. In contrast, electric field $E$ affects the selection of spin chirality\cite{CuFeO2_PolarizedNeutron}. If $E$ is applied along [110], three multiferroic domains are selected as depicted in Fig. 1(k). Further application of $H$ should sort domains; a domain distribution as illustrated in Figs. 1(l) or (m) is favored with {\Hin} or {\Hbar}. Comparing these arrangements (Figs. 1(k)-(m)), {\Pin} should increase with {\Hin} and decrease with {\Hbar}. This idea well explains the observed results, and similar effects have recently been reported for multiferroic delafossite CuCrO$_2$\cite{CuCrO2_Seki, CuCrO2_Kimura}. With {\Hin}- and {\Hbar}-scan at 2K, both $M$ and {\Pin} show anomalies around 5T only in the field-increasing run (Figs. 2(e) and (f)), suggesting the field-irreversible rearrangement of domain distribution at this specific field direction.

\begin{figure}
\begin{center}
\includegraphics*[width=8.5cm]{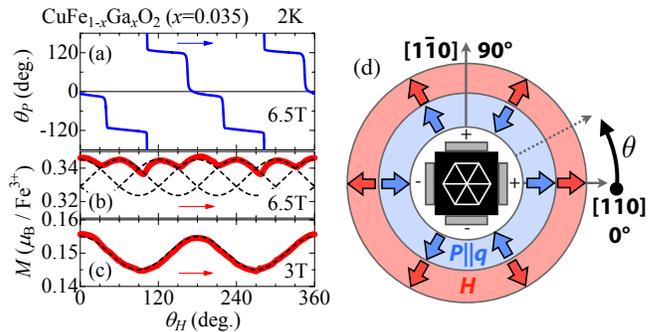}
\caption{(color online). (a) Relationship between the directions of $P$ and $H$, both confined within the (001) plane. $\theta_P$ ($\theta_H$) denotes the angle between $P$- ($H$-) direction and the [110] axis. (b),(c) Corresponding variations of $M$. Dashed lines indicate theoretically expected behaviors. (d) Relationship among $P, q$ and $H$ at $\theta_H = (60n)^\circ$ ($n$: integer).} 
\end{center}
\end{figure}

Next, we have investigated the vector components of $P$ in response to the in-plane $H$ rotating around the $c$-axis. For this purpose, the both [110] and [1$\bar{1}$0] components of $P$ were measured simultaneously with two pairs of electrodes. In this configuration, both $P$ and $H$ can be expressed as the two-dimensional vector on the (001) plane. Hereafter, we define the angle between $H$ ($P$) and the [110]-axis as {\AngleH} ({\AngleP}). Since the specimen was cooled with $H$ and $E$ both applied along [110], we assume the uniform initial domain state as shown in Fig. 1(l). 

Figures 3(a) and (d) show {\Pin} and {\Pbar} as a function of {\AngleH}, measured at $H=6.5$T without $E$. Both {\Pin} and {\Pbar} show a periodic change with the cycle of 180$^\circ$. To see the development of $P$ more directly, we plot the obtained {\AngleP} against {\AngleH} for the {\AngleH}-increasing run (Fig. 4(a)). In agreement with the expected initial state in Fig. 1(l), the relationship $P \parallel H \parallel [110]$ is confirmed at ${\theta_H}=0$. As {\AngleH} increases, $P$ suddenly flops by about 120$^\circ$ at $\theta_H=41^\circ$, and the relation that $-P \parallel H \parallel [100]$ holds at $\theta_H=60^\circ$. Since $H$ favors multiferroic domains with $q \parallel P \parallel \pm H$, this transition can be considered as the flop of magnetic $q$-vector from $q \parallel [110]$ to $q \parallel [100]$. Note that both transitions to $P \parallel [100]$ and $P \parallel - [100]$ states seem possible from the $P \parallel [110]$ initial state, but in reality only the $P \parallel - [100]$ state is selected here. Such a $q$-flop (or $120^\circ$-flop of $P$)  is observed upon every $60^\circ$-rotation of $H$, consistent with the symmetry of underlying triangular lattice. The absolute value $|P|$ (not shown) is almost constant for any {\AngleH}. 

Between the two opposite directions of $H$-rotation, a relatively large hysteresis is found in $P$. If we estimate the critical {\AngleH} from the average of both direction of scans, the $q$-flop is always centered at $\theta_H = (30 +60n)^\circ$ ($n$: integer). This agrees with the equilibrium point of two magnetic $q$-vectors. The appearance of hysteresis means that an excess gain of Zeeman energy is needed to overcome the potential barrier height. When $H$ is reduced from 6.5T to 4.5T, the reduction of Zeeman energy leads to expansion of the hysteresis (Figs. 3(b) and (e)). Below 3.5T, the potential barrier cannot be overcome and no $q$-flop behavior is observed (Figs. 3(c) and (f)). 

Corresponding variation of $M$ as a function of $\theta_H$ is shown in Figs. 4 (b) and (c). To obtain the $q \parallel [110]$ initial state, the specimen was cooled at $\theta_H=0^{\circ}$ with $H=6.5$T and then magnitude of $H$ is fixed prior to measurements. Without $q$-flop, this initial state should give $\Delta M \propto (\chi_{\parallel} - \chi_{\perp}) \cos (2\theta_H)$, where $\chi_{\parallel}$ and $\chi_{\parallel}$ denotes $\chi$ parallel or perpendicular to $q$. This agrees well with $\theta_H$ dependence of $M$ at 3T (Fig. 4(c)), indicating the robustness of single-$q$ state and the non-volatile nature of $q$-domain distribution. In contrast, the profile at 6.5T has a period of $60^{\circ}$ and corresponding sinusoidal curve is shifted by $60^{\circ}$ upon every $P$-flop transition. This confirms the emergence of $q$-flop as the origin of $P$-flop and the clamping of ferroelectric and magnetic domain walls.

Figure 4(d) illustrates the relationship between $P$ and $H$ at $\theta_H$=$(60n)^\circ$. At each {\AngleH}, we could confirm $P \parallel \pm H$, in agreement with the $q$-flop model. Importantly, upon each transition, $P \parallel H$ and $P \parallel -H$ alternately appears. Generally, magnetic domains can be mutually converted by symmetry operation that is broken by magnetic order\cite{Schmid}. If we apply space inversion to Fig. 1(g), a domain with opposite $P$ and reversed spin chirality can be obtained (Fig. 1(h)). Mirror operation on Fig. 1(g) generates another domain with reversed spin chirality (Fig. 1(i)). Likewise, we can reproduce all six $P$-domains and determine their corresponding spin chirality (Fig. 1(j)). From this relationship, it is concluded that the chirality of spin spiral is always conserved upon the $q$-flop. 

Recently, a similar systematic behavior of spin chirality has been observed in some ferroelectric helimagnets: Upon $90^\circ$-flop of $q$ under rotating $H$ on ZnCr$_2$Se$_4$, the spin chirality is preserved in lower-$H$ region but reversed in higher-$H$ region\cite{ZnCr2Se4}. In case of $90^\circ$-flop of spin spiral plane (with fixed $q$) in unidirectional $H$ on MnWO$_4$, spin chirality after the transition can be selected by slight tilt of $H$\cite{Chiral_MnWO4}. Since two opposite chiral states are energetically degenerated under $H$, selection of odd chirality upon magnetic transition was explained by the energy difference between two possible domain wall (DW) structures connecting domains with the same or opposite chirality. The behavior of spin chirality depends on the given $H$-condition, therefore the stability of DW should be modified by $H$. This situation likely occurs in {\CFGO}: When rotating $H$ induces the $P$-flop from $P_1$ to $P_2$, the angle between $P_1$ and $P_2$ across the DW becomes $120^\circ$ for the same chirality (120$^\circ$-DW), and $60^\circ$ for the opposite chirality (60$^\circ$-DW). The present robustness of spin chirality can be explained, only provided that 120$^\circ$-DW is more stable than the 60$^\circ$-DW. Detailed theoretical calculation of relative stability of two types of DWs considering all magnetic, dielectric, structural, and chiral degree of freedoms, as well as direct observation of domain dynamics, is highly desirable.

In summary, we demonstrate the persistent magnetic control of six ferroelectric domains for triangular-lattice ferroelectric helimagnet {\CFGO} ($x$=0.035). The flop of the magnetic modulation vector is induced by every 60$^\circ$-rotation of in-plane $H$ around the $c$-axis, which leads to every 120$^\circ$-flop of $P$-direction within the triangular-lattice basal plane. The chirality of spin-spiral is always conserved upon the $P$-flop, which may reflect the stability of the specific multiferroic domain wall structure. In more general, the nature of multiferroic DW in a ferroelectric helimagnet should play a key role in determining the $P$-direction or spin-chirality upon the $H$-induced $q$-flop; this may enable the unusual selective $P$-domain switching with varying direction and magnitude of $H$. Ferroelectric helimagnets with highly symmetric crystal structure like cubic, tetragonal or hexagonal lattices can host the multiple $H$-switchable $P$-domains as revealed in this study, and hence are promising for ME control with non-volatility and multiple-valued nature.

The authors thank T. Arima, H. Ito, Y. Yamasaki, Y. Kohara, S. Ishiwata, S. Tanaka, and H. Katsura for enlightening discussions. This work was partly supported by Grants-In-Aid for Scientific Research (Grant No. 16076205, 17340104) from the MEXT of Japan.

\end{document}